\def\bclbig{\big\langle\!\!\big\langle}
\def\bcrbig{\big\rangle\!\!\big\rangle}
\def\bcl{\langle\!\langle}
\def\bcr{\rangle\!\rangle}

\documentstyle[eqsecnum,prd,aps,epsfig]{revtex}
\begin{document}
\preprint{
\vbox{
\halign{&##\hfil\cr
	& TRI-PP-97-32\cr}}
}
\title{Comment Concerning Relativistic Corrections in NRQCD }
\author{Ivan Maksymyk}
\address{TRIUMF, 4004 Wesbrook Mall, Vancouver, Canada, V6T 2A3}
\maketitle
\begin{abstract}
We examine the subtle differences between two possible frameworks
for the calculation of quarkonium production. The differences
between the two methods are not a concern when one calculates
only to leading order in the relativistic expansion, but when
relativistic corrections are included, the two formulations
seem at first to be at variance with one another.
They can be reconciled however via a relativistically corrected
mass relation relating the boundstate mass and the quark mass.
\end{abstract}

\section{Introduction}

We present here an observation regarding relativistic
corrections in factorization formulas for
quarkonium production, as calculated in the non-relativistic
QCD framework \cite{bbl} .   We point out that one can conceive of two subtlely
different frameworks in which calculations
of quarkonium production might be
carried out.
We illustrate the algorithm for
both approaches through
the detailed presentation of a simple example.
We explain that the results of the methods differ
on two points: on the question
of the definitions of the NRQCD matrix elements;
and on the question
of the mass parameters that appear in NRQCD factorization
formulas.  In Method I,
factorization formulas are computed in such a manner that
only the quark mass $m_c$ appears, while in Method II,
the formulation generally entails the mass of the quarkonium
boundstate (which we will denote with the upper case $M_H$)
as well as $m_c$.

The difference between the two methods
is not a concern when one calculates
quarkonium production and decay rates only to leading order
in the relativistic expansion.  In that approximation,
one sets $M_H = 2m_c$, and the two methods give compatible results.
However, when
relativistic corrections are included
(and when $M_H$ no longer is equatable to $2m_c$),
we find that the
two formulations seem at first to be at variance with one another.

The results of the two methods can be reconciled
using a {\it relativistically corrected mass relation},
which is an equation relating $m_c$, $M_H$ and
the NRQCD production matrix elements.

A variant of the
relativistically corrected mass relation has recently been discussed
in Ref.~\cite{ctone}.  The appearance
of this work increases the possible interest of
the NRQCD community in our observation.

In order to illustrate our observation, we take the simplest possible
example, that of
charmonium production from the collision of two
a ficticious elementary (colorless)
scalars, denoted by $\phi$:
$$
\phi + \phi \rightarrow {\mbox{charmonium}}.
$$
The underlying short-distance physics is
the partonic process
%
$$
\phi + \phi \rightarrow  c + \overline{c}
$$
%
which is
described by the
interaction term
\begin{equation}
\label{interaction}
{\cal L}_{int} \,\, = \,\, g \,
\phi(x)\, \phi(x) \,\, \overline{\Psi}_i\!(x)\gamma_5\Psi^i\!(x)
\end{equation}
where $\Psi^i\!(x)$ is the four-spinor field operator for the charm quark
(with color index $i$),
and where $g$ is a coupling constant of mass dimension $-1$.
The reaction $\phi + \phi \rightarrow c + \overline{c}$ produces
a $c\overline{c}$ pair in a color-singlet state with
angular momentum ${}^{2S+1}L_J = {}^1S_0$.
This $c\overline{c}$ can hadronize into the $\eta_c$ charmonium particle.
We choose this process because of the simplicity of the spin
structure and because there are no color-octet contributions
which would distract from our main point.

The determination of production factorization formulas in NRQCD
is always based, in one way or another,
on the Feynman amplitude ${\cal M}$
for the production of a {\it free} heavy quark and {\it free} heavy
antiquark.
In particular, we desire to know
how this quantity depends on ${\bf q}$,
the relative three momentum of the quark and antiquark in the
quark-antiquark restframe.

In our toy example,
the Lorentz-invariant Feynman amplitude is given by
\begin{equation}
\label{amplitude}
{\cal M}
\big( c\overline{c}({\bf q}) \big) \; = \; g\, \overline{u}v
\; = \; 2 g \,\, E_q \,\, \xi^\dagger_\sigma \zeta_\tau .
\end{equation}
(Colour indices have been suppressed.)
The Dirac four-spinors $u$ and $v$ are normalized so that
$\bar{u}u = \bar{v}v = 2m_c$.  The passage from the
middle piece
of Eq.~\ref{amplitude}
to the right-hand side is an exact relation, and not a
nonrelativistic approximation.
It entails recasting the four-spinors (from the Pauli-Dirac
representation) in terms of the two-spinors $\xi$ and $\zeta$,
as described in Ref.~\cite{bc}.  $\xi$ is the two-spinor for
the heavy quark, and $\zeta$ (called $\eta$ in Ref.~\cite{bc})
is the two-spinor for the heavy
antiquark.
$\sigma$ and $\tau$ label heavy quark spins;
they are not spinor component indices.
The two-spinors are normalized so that
$\xi_\sigma^\dagger\xi_\tau = \delta_{\sigma\tau}$.
As to $E_q$, this is
simply a short form
for $\sqrt{{\bf q}^2 + m_c^2}$.  The invariant mass of
the system consisting of the pair of free on-shell heavy quarks
is $2E_q$.

The spinor-structure given in Eq.~\ref{amplitude}
is such that multiplying the expression by the
Clebsch-Gordon coefficients $\langle 1/2, \sigma; 1/2, \tau | 0,0 \rangle$
and then summing over spins gives a non-zero result. However, doing
the same with the Clebsch-Gordon coefficients
$\langle 1/2, \sigma; 1/2, \tau | 1,M \rangle$
gives zero.  We therefore see that Eq.~\ref{amplitude} represents
the production of a $c\overline{c}$ pair in a spin state
$J,M_J = 0,0$, as was previously stated.

\section{Relativistically Corrected Factorization Formulas: Method I}

\subsection{The Concept of Matching}

The NRQCD factorization formalism exploits the equivalence of
perturbative QCD and perturbative NRQCD.  The computation of
the factorization formula for the production of
a quarkonium boundstate starts
with a computation of the production rate for free on-shell
heavy quarks and antiquarks;
this computation is performed in some perturbative model
of underlying short-distance physics, which could
be perturbative QCD, electroweak physics, or (as here) a toy model.
In a separate calculation, one uses NRQCD
perturbation theory
to compute the free-quark matrix elements of the four-fermion
operators of the NRQCD effective theory.
The perturbative equivalence of NRQCD
and the full underlying theory
is expressed by the matching condition
(Eq.~(6.7) of Ref.~\cite{bbl})
\begin{equation}
\label{bblmatchingcondo}
\sigma \big( c\overline{c}({\bf q}) \big) \Big|_{{\mbox{pert}}}
= \sum_n
\frac{F_n(m_c)}{m_c^{d_n - 4}}
\,
\big\langle 0 | {\cal O}^{c\overline{c}(q)}_n | 0 \big\rangle
\Big|{}_{{\mbox{pert NRQCD }}\;\; . }
\end{equation}
The left-hand side
is to be computed in the perturbative model of underlying physics.  It
is the rate
for the production of a heavy quark pair,
written expressly as a function
of ${\bf q}$, the relative three-momentum of the
$c$ and $\overline{c}$ in the  $c\overline{c}$ restframe.
As well as being a function of ${\bf q}$,
the left-hand side is also a
function of whatever other kinematic variables are
required for parametrizing the process at hand.
As to the right-hand side, it is simply a linear combination
NRQCD free-quark matrix elements
$\langle 0 | {\cal O}^{c\overline{c}(q)}_n | 0 \rangle$
such as those defined
in Appendix I.  The sum over $n$ in the right-hand side is
a Taylor series;  each of the matrix elements
is a function of ${\bf q}$ and $m_c$ only, with the power
in ${\bf q}^2/m_c^2$ increasing with each higher term.
The matching condition is intended to allow a determination
of the short-distance coefficients $F_n$.  By definition,
these are independent of ${\bf q}$.  How the matching condition
is to be employed will become clear in the example below.
The basic idea is that, by calculating the production rate
in the full underlying theory on the left-hand side, and by massaging it into
a Taylor series of the form of the right-hand side, one can then
read off the $F_n$.
$d_n$ is the mass dimension of the
product of the field operators in ${\cal O}_n$.

Once the short-distance coefficients $F_n$ have been calculated
for free on-shell heavy quarks in the perturbative theory, the
rate for the production of some quarkonium boundstate $H$
is then known to be (Eq.~(6.4) of Ref.~\cite{bbl})
\begin{equation}
\label{boundstateffbbl}
\sigma ( H )
= \sum_n
\frac{F_n(m_c)}{m_c^{d_n - 4}}
\,
\big\langle 0 | {\cal O}^{H}_n | 0 \big\rangle\; .
\end{equation}
The above expression is a factorization formula.
The NRQCD non-perturbative matrix elements
$\big\langle 0 | {\cal O}^{H}_n | 0 \big\rangle $
are to be thought of as empirical parameters.
A fundamental feature of the NRQCD factorization formalism is that
the coefficients $F_n$ are universal to both free quark and
boundstate production.
In general, the short-distance coefficients and the hadronic
matrix elements also depend on the NRQCD cut-off, but we will
not bother to display this dependence, since it is not a point
of focus in our discussion.

\subsection{Manipulation of Phase Space for Method I}

In Method I, we interpret the matching condition, Eq.~\ref{bblmatchingcondo},
to involve (on both sides of the equation) the rate of production of
free heavy quarks; this interpretation defines Method I.
In such an approach,
the physical meaning of
$\sigma\big( c\overline{c}({\bf q}) \big)$
is fixed via the concept of the total cross-section for free on-shell $c\overline{c}$ production:
\begin{equation}
\label{essenceofmethodone}
\sigma\big( c\overline{c} \big)  \; = \;
\int \frac{d^3{\bf q}}{(2\pi)^3} \;\sigma\big( c\overline{c}({\bf q}) \big) \;\; .
\end{equation}
By construction, $\sigma\big( c\overline{c}({\bf q}) \big)$
depends on ${\bf q}$, $m_c$ and
other kinematic variables of the free $c\overline{c}$
production process,
but it logically does not depend on the mass of quarkonium boundstate, $M_H$.

Our goal in this section is
to obtain a useful general expression for
$ \sigma\big( c\overline{c}({\bf q}) \big) $ for the purposes of
matching in Method I.
We procede by first computing the production rate for a pair of
free on-shell quarks.  The total rate is
given by the well-known formula
\begin{equation}
\label{basicrateformula}
\sigma(c\overline{c})
=
\int \;
\frac{d^3 {\bf p}_c}{(2\pi)^3 2 p_c^0}\;
\frac{d^3 {\bf p}_{\bar{c}} }{(2\pi)^3 2 p_{\bar{c}}^0}\;
\sum_X \;
\frac{1}{{\mbox{flux}}}\; \delta^4(p_i - X - p_c - p_{\bar{c}}) \;
(2 \pi)^4 \;
|\overline{{\cal M}
\big(  c\bar{c}({\bf q}) \big)
}|^2
\end{equation}
where, since we are only considering spin-summed rates,
we sum over all quantum numbers of the
final state free heavy quark and free heavy antiquark.
Here,
$X$ represents the sum of the momenta of all final state particles
other than the two heavy quarks.
$\sum_X$ represents the measure of phase space integration over
all such momenta.

Let us now define the total and relative four-momenta of the
heavy quarks:
\begin{eqnarray}
\label{predef}
P & \equiv  & p_c + p_{\bar{c}} \nonumber\\
\label{qredef}
2 Q & \equiv & p_c - p_{\bar{c}} .  \nonumber
\end{eqnarray}
$P$ is the total four-momentum of the $c\bar{c}$ system
in the lab frame.
$Q$ is the relative four-momentum of the $c$ and $\bar{c}$
in the lab frame.
The lab-frame four-momentum $Q$ is related to the
rest-frame three-momentum ${\bf q}$ by the relation
$$
Q^\mu = \Lambda^\mu_{\,\,i} q^i
$$
where $\Lambda^\mu_{\,\, \nu}$ is the Lorentz boost matrix connecting
the lab frame with the $c\overline{c}$ restframe.
Recalling that the invariant mass of the $c\overline{c}$
system is $2E_q \equiv 2 \sqrt{m_c^2 + {\bf q}^2}$, we see that
the components of the total four-momentum of
the $c\bar{c}$ system are
%
$$
P =
\Big\{ \sqrt{4E_q^2 + {\bf P}^2}, {\bf P} \Big\}
$$

We now manipulate the phase space integral in
Eq.~\ref{basicrateformula} by defining the new variables
of integration
\begin{eqnarray}
\label{predefint}
{\bf P} & = & {\bf p}_c + {\bf p}_{\bar{c}} \nonumber\\
\label{qredefint}
2 {\bf Q} & = & {\bf p}_c - {\bf p}_{\bar{c}} . \nonumber
\end{eqnarray}
With these new variables,
the total rate is written
$$
\sigma \big( c\bar{c} \big)
=
\int
\;
\frac{d^3 {\bf Q} \;  d^3 {\bf P} }
{ (2 \pi)^6 \; 2p_c^0 \; 2p_{\overline{c}}^0}\; \sum_X
\;
\frac{1}{{\hbox{flux}}}
\delta^4(p_i - X - P ) \; (2\pi)^4\;
|\overline{{\cal M}
\big( c\bar{c}({\bf q}) \big)
}|^2 \;\; .
$$
The Jacobian for the above variable redefinition is unity.

We now change variables of integration once more, replacing
$Q^i$ in favor of $q^j$.  Concerning this transformation, we have the
following exact identity:
\begin{eqnarray}
&&\int \; d^3 {\bf Q}
\;
\Bigg[ \frac{1}{2 p_c^o \; 2 p_{\overline{c}}^0} \Bigg] \nonumber\\
& = & \int \Bigg[ d^3 {\bf q} \; \Big| {\hbox{Det}}
\frac{dQ^i}{dq^j}
\Big| \Bigg]
\;
\Bigg[ \frac{1}{2 p_c^o \; 2 p_{\overline{c}}^0} \Bigg] \nonumber\\
& = &
\int \Bigg[ d^3 {\bf q}\frac{P^0}{2 E_q}
\left(
 1 - \frac{{\bf P}^2 {\bf q}^2}{12 (P^0)^2 E_q^2}
 \right) \Bigg]
\;
\Bigg[ \frac{1}{(P^0)^2}
\left( 1 - \frac{{\bf P}^2 {\bf q}^2}{12 (P^0)^2 E_q^2} \right)^{-1}
\Bigg] \;\; , \nonumber
\end{eqnarray}
so that the total rate can be written in the
form:
%
$$
\sigma \big( c\bar{c} \big) =
\int
\;
\frac{d^3 {\bf q}}{(2\pi)^3} \; \frac{ d^3 {\bf P} }
{ (2 \pi)^3 2E_q P^0}\; \sum_X
\;
\frac{1}{{\mbox{flux}}}
\; \delta^4(p_i - X - P) \;
(2 \pi)^4\;
|\overline{{\cal M}
\big(  c\bar{c}({\bf q}) \big)
}|^2 \;\; .
$$
We are now able to extract $\sigma\big( c\overline{c}({\bf q}) \big)$,
the rate as a function of ${\bf q}$.
This is seen to be
\begin{equation}
\label{rateq}
\sigma \big( c\bar{c}({\bf q})  \big) =
\int
\;
\frac{ d^3 {\bf P} }
{ (2 \pi)^3 2E_q P^0}\; \sum_X
\;
\frac{1}{{\mbox{flux}}}
\; \delta^4(p_i - X - P) \; (2\pi)^4 \;
|\overline{{\cal M}
\big(  c\bar{c}({\bf q}) \big)
}|^2 \;\; .
\end{equation}
\subsection{Matching Condition for Method I}
Eq.~\ref{rateq} gives an expression for the
left-hand side of the basic matching condition, Eq.~\ref{bblmatchingcondo}.
The matching condition now becomes
\begin{equation}
\label{bblfinal}
\int \;
\frac{ d^3 {\bf P} }
{ (2 \pi)^3 2E_q P^0}\; \sum_X
\;
\frac{1}{{\mbox{flux}}}
\; \delta^4(p_i - X - P) \; (2\pi)^4 \;
|\overline{{\cal M}
\big(  c\bar{c}({\bf q}) \big)
}|^2  =
\frac{F_n(m_c)}{m^{d_n - 4}}
\big\langle 0 | {\cal O}_n^{c\overline{c}(q)}|0 \big\rangle\;\; .
\end{equation}
We will take this as our matching condition for Method I.
It is to be used as follows:  Keeping in mind that
$E_q$ and $P^0$ are functions of ${\bf q}$,
one computes the left-hand side in the
underlying perturbative theory; one expresses this as
a Taylor series in ${\bf q}^2/m_c^2$, and then massages it
into the form of the right-hand side; one then reads off the $F_n$.
Because of the universality of the $F_n$, the $F_n$ determined
in the matching procedure
also serve in the factorization formula for the production of
boundstate quarkonium, Eq.~\ref{boundstateffbbl}, which is
%
$$
\sigma ( H )
= \sum_n
\frac{F_n(m_c)}{m_c^{d_n - 4}}
\,
\big\langle 0 | {\cal O}^{H}_n | 0 \big\rangle\; .
$$
%

In the next section, we present an illustrative example of the method.

\subsection{Example of Calculation in Method I}

We now consider our toy example,
$\phi + \phi \rightarrow \eta_c$.
The underlying interaction is expressed in Eq.~\ref{interaction}.
Let us compute the left-hand side of the matching condition
Eq.~\ref{bblfinal}.  In this example, there is no sum $\sum_X$.
We have
\begin{eqnarray}
\sigma \big( c\overline{c}({\bf q}) \big)
 & = & \int
\frac{ d^3 {\bf P} }
{ (2 \pi)^3 2E_q P^0}\; \frac{1}{{\mbox{flux}}} \;
\;
\; \delta^3 ( {\bf P})  \delta(E_f - E_i)
(2 \pi)^4 \;
|\overline{{\cal M}
\big(  c\bar{c}({\bf q}) \big)
}|^2\nonumber\\
& = & %
 \int
\frac{ d^3 {\bf P} }
{ (2 \pi)^3 (2E_q)^2 }\; \frac{1}{8 E_q^2} \;
\;
\; \delta^3 ( {\bf P})  \delta(E_f - E_i)
(2 \pi)^4 \;
|\overline{{\cal M}
\big(  c\bar{c}({\bf q}) \big)
}|^2\nonumber\\
& = & (2 \pi) \frac{1}{32 E_q^4} \;\delta(E_f - E_i) \;
|\overline{{\cal M}
\big(  c\bar{c}({\bf q}) \big)
}|^2\nonumber\\
& = & (2 \pi) \frac{1}{32 E_q^4} \;\delta(E_f - E_i) \;
4 E_q^2 g^2 N_c \; \sum_{\sigma\tau}\;
\zeta^\dagger_\tau\xi_\sigma \; \xi_\sigma^\dagger \zeta_\tau
\nonumber\\
& = & \frac{ \pi g^2}{4 m_c^2} \; \delta(E_f - E_i) \;
\Bigg[ 1 - \frac{{\bf q}^2}{m_c^2} + \cdots \Bigg]
\;
N_c \; \sum_{\sigma\tau}\;
\zeta^\dagger_\tau\xi_\sigma \; \xi_\sigma^\dagger \zeta_\tau
\end{eqnarray}
where we have used the Feynman amplitude given in Eq.~\ref{amplitude}.
This can be written in terms of the free-quark NRQCD matrix elements
given in Appendix I:
\begin{equation}
\label{cleverone}
{\mbox{lhs}}
 =
\frac{ \pi g^2}{4 m_c^2} \; \delta(E_f - E_i) \;
\Bigg[
\big\langle 0 | {\cal O}_1^{c\overline{c}(q)}({}^1S_0) | 0 \big\rangle
-
\frac{1}{ m_c^2}
\big\langle 0 | {\cal P}_1^{c\overline{c}(q)}({}^1S_0) | 0 \big\rangle
+ \cdots \Bigg]
\end{equation}
We infer from this that
the production rate for the charmonium boundstate $\eta$
is given by
\begin{equation}
\label{happyone}
\sigma(\eta)
 =
\frac{ \pi g^2}{4 m_c^2} \; \delta(E_f - E_i) \;
\Bigg[
\big\langle 0 | {\cal O}_1^{\eta}({}^1S_0) | 0 \big\rangle
-
\frac{1}{ m_c^2}
\big\langle 0 | {\cal P}_1^{\eta}({}^1S_0) | 0 \big\rangle
+ \cdots \Bigg]
\end{equation}
This is the factorization formula for Method I.
It corresponds to Eq.~(6.8a)
of Ref.~\cite{bbl}.
We have included the leading piece and the first
relativistic correction.

We must reemphasize that the factorization
formulas in this approach make no explicit reference to the boundstate
mass $M_H$.  The short-distance coefficients can logically
depend only on $m_c$.

\section{Relativistically Corrected Factorization Formulas: Method II}

\subsection{An ``Effective Amplitude Squared'' for Quarkonium Production}

Another approach to carrying out calculations of quarkonium production
rates is suggested in Ref.~\cite{bc}.
It is our interpretation that in this work, it is implicitly assumed
that there exists a meaningful theoretical quantity
 --- an ``effective Feynman amplitude
(squared), ''
$|{\cal M}(H)|^2$ ---  for the production of a quarkonium boundstate.
This assumption is tacitly present in the first equation in Ref.~\cite{bc},
which gives the cross-section for quarkonium
boundstate production in terms of $|{\cal M}(H)|^2$:
\begin{equation}
\sigma(H)  =
\int
\frac{d^3 {\bf P}_H}{(2\pi)^32E_H}  \sum_X \;\frac{1}{{\hbox{flux}}}\
(2\pi)^4 \delta^4(p_i - X - P_H) \;
|\overline{{\cal M}(H)}|^2 \;\; .
\end{equation}
Here,
$X$ represents the sum of the momenta of all final state particles
other than the quarkonium boundstate.
$\sum_X$ represents the measure of phase space integration over
all such momenta.
It is important to appreciate that here
$P_H$ represents the four-momentum of the quarkonium boundstate.
The components of $P_H$ are
$\{ E_H, {\bf P}_H \}$,
where $E_H$ is defined as $\sqrt{M_H^2 + {\bf P}^2_H}$ and where
$M_H$ is the mass of the boundstate.

Ref.~\cite{bc} postulates that the production rate
for quarkonia can be
written in a factorized form
\begin{equation}
\label{bcboundstate}
\sigma(H)  =
\int
\frac{d^3 {\bf P}_H}{(2\pi)^32E_H} \;\frac{1}{{\hbox{flux}}} \;
\sum_n \frac{ {\cal F}_n(m_c, {\bf P}_H) }{m_c^{d_n - 4}} \;
\bclbig  0 |
{\cal O}_n^H
| 0 \bcrbig \;\; .
\end{equation}
We use the symbols $\bcl$ and $\bcr$ for the bras and kets of
Method II as a reminder that
here the matrix elements
$\bcl 0 |{\cal O}_n^H| 0 \bcr $ are defined
differently from the
$ \big\langle 0 |
{\cal O}_n^H | 0 \big\rangle$ of Method II.
The reader must examine Appendices I and II for the details concerning
the different conventions and definitions.
The distinction concerns the normalization of states.
The ${\cal F}_n(m_c, {\bf P}_H)$ are the short-distance coefficients
of Method II, distinct from the $F_n$ of Method I.
With the states being normalized differently, the NRQCD matrix elements
(and also the short-distance coefficients) are of different
mass dimension in the two methods.

\subsection{Matching Condition for Method II}

In order to determine the short-distance coefficients ${\cal F}_n$
for a given quarkonium production process, we require a matching
condition, analagous to Eq.~\ref{bblfinal}.  The condition
proposed in Refs.~\cite{bc} and \cite{eric} is
\begin{equation}
\label{bcmatchingcondo}
\sum_X
\; (2 \pi)^4 \;
\delta^4 \big( p_i - X- P({\bf q}) \big)
\;
|\overline{{\cal M}\big( c\bar{c}({\bf q}) \big)} |^2  =
\sum_n \frac{{\cal F}_n(m_c, {\bf P}) }{m_c^{d_n - 4}} \;
\bclbig 0 |
{\cal O}_n^{c\bar{c}(q)}
| 0 \bcrbig
\end{equation}
This matching condition lends itself to
being interpreted
in the following manner:
one computes $|{\cal M}\big( c\overline{c}({\bf q}) \big)|^2$
in the underlying perturbative theory;
this allows one to express the complete left-hand side as
a Taylor series in ${\bf q}^2/m_c^2$ (it must
be kept in mind that $P^0$ depends on ${\bf q}$); one then massages
the resulting expression
into the form of the right-hand side; finally one
reads off the ${\cal F}_n$.
These ${\cal F}_n$ serve in the factorization formula for the production of
boundstate quarkonium, Eq.~\ref{bcboundstate}.

In the next section, we present an illustrative example of the method.

\subsection{Example of Calculation in Method II}

We now calculate the factorization formula for
our toy example using Method II.
The left-hand side of the matching condition,
Eq.~\ref{bcmatchingcondo}, is
\begin{eqnarray}
{\mbox{lhs}} & = &
\delta^3({\bf P}) \delta(E_f - E_i) \; (2\pi)^4\;
|\overline{{\cal M}
\big(  c\bar{c}({\bf q}) \big)
}|^2\nonumber\\
& = &
\delta^3({\bf P}) \delta(E_f - E_i) \; (2\pi)^4\;
4 g^2 E_q^2 N_c \; \sum_{\sigma\tau}\;
\zeta^\dagger_\tau\xi_\sigma \; \xi_\sigma^\dagger \zeta_\tau
\end{eqnarray}
This can be  written in terms of the free-quark NRQCD matrix elements
given in Appendix II:
\begin{equation}
\label{clevertwo}
{\mbox{lhs}}
=
g^2 \delta^3({\bf P}) \delta(E_f - E_i) \; (2\pi)^4\;
\bclbig 0 | {\cal O}_1^{c\overline{c}(q)}({}^1S_0) | 0 \bcrbig
\end{equation}
Interestingly, the Taylor expansion stops after the first
term, but this is not a generic feature of this method.  In general,
there would be an infinite series of terms in powers of
${\bf q}^2/m_c^2$.
We infer from Eq.~\ref{clevertwo} that
the production rate for the charmonium boundstate $\eta_c$
is given by
\begin{eqnarray}
\label{happytwo}
\sigma(\eta)
&  = & \int \frac{d^3 {\bf P_H}}{(2\pi)^3 2 E_H} \;\;\;
g^2 \delta^3({\bf P_H}) \delta(E_f - E_i) \; (2\pi)^4\;\;
\bclbig 0 | {\cal O}_1^\eta ({}^1S_0) | 0 \bcrbig \nonumber\\
& = & \frac{\pi g^2 }{ 2 M_\eta^3}
\bclbig 0 | {\cal O}_1^\eta ({}^1S_0) | 0 \bcrbig   \delta(E_f - E_i)
\end{eqnarray}
This is the factorization formula for Method II.

\section{Comparison of the Two Methods}

It is instructive to gather together the matching conditions for
the two methods, for the purposes of comparison.
\begin{eqnarray}
\int
\frac{ d^3 {\bf P} }
{ (2 \pi)^3 2E_q P^0}\; \sum_X \; \delta^4(p_i  -  X  - P({\bf q}) ) \;
\;
\frac{1}{{\mbox{flux}}}
\; (2\pi)^4 \;
|\overline{{\cal M}
\big(  c\bar{c}({\bf q}) \big)
}|^2  & \; =  \; &
\sum_n
\frac{F_n(m_c) }{m_c^{d_n - 4}}
\big\langle 0 | {\cal O}_n^{c\overline{c}(q)}|0 \big\rangle \nonumber\\
\;
\sum_X \; \delta^4(p_i - X - P({\bf q})) \;\;
\;
\frac{1}{{\hbox{flux}}}\; (2 \pi)^4 \;
|\overline{{\cal M}\big( c\bar{c}({\bf q}) \big)} |^2  & \; = \; &
\sum_n \frac{{\cal F}_n (m_c, {\bf P})  }{m_c^{d_n - 4}}
\bclbig 0 |
{\cal O}_n^{c\bar{c}(q)}
| 0 \bcrbig
\end{eqnarray}
where in both cases, the four-momentum
$P({\bf q})$ is a function of ${\bf q}$ in the
sense that, while ${\bf P}$ is strictly independent of ${\bf q}$,
$P^0({\bf q})$ is given by $P^0({\bf q}) =
\sqrt{4E_q + {\bf P}^2}$.
This last point is not an issue in the toy model, but is an
issue for the more general situation in which $p_i^0$
(the total initial energy) is fixed
(and is independent of ${\bf q}$), as in the case of
$b \rightarrow J/\psi + s$.

Once the short-distance coefficients have been found
with these matching conditions, the total production
rate is given by
\begin{eqnarray}
\label{useone}
\sigma ( H )
& = & \sum_n
\frac{F_n (m_c) }{m_c^{d_n - 4}}
\,
\big\langle 0 | {\cal O}^{H}_n | 0 \big\rangle\; \nonumber\\
\label{usetwo}
\sigma(H)  & = &
\int
\frac{d^3 {\bf P}_H}{(2\pi)^32E_H} \;\frac{1}{{\hbox{flux}}}
\;
\sum_n \frac{ {\cal F}_n(m_c, {\bf P}_H) }{m_c^{d_n - 4}} \;
\bclbig  0 |
{\cal O}_n^H
| 0 \bcrbig \nonumber\\
& = &
\sum_n \frac{ {\cal F}'{}_n(m_c, M_H) }{m_c^{d_n - 4}} \;
\bclbig  0 |
{\cal O}_n^H
| 0 \bcrbig
\end{eqnarray}

The reader will appreciate, by trying a few simple examples, that
the state normalizations and conventions in Appendix I are ideally
suited to Method I, and cannot be naturally employed in Method II.
The state normalizations and conventions in Appendix II are ideally
suited to Method II, and cannot be naturally employed in Method I.

\section{Reconciliation of the Two Methods ? }

We have obtained the following factorization formulas
for the charmonium production process
$ \phi + \phi \rightarrow \eta $:
\begin{eqnarray}
\label{toyresults}
{\mbox{Method I:}} \;\;\;\; & \sigma(\eta) & =
\frac{ \pi g^2}{4 m_c^2} \; \delta(E_f - E_i) \;
\Bigg[
\big\langle 0 | {\cal O}_1^{\eta}({}^1S_0) | 0 \big\rangle
-
\frac{1}{ m_c^2}
\big\langle 0 | {\cal P}_1^{\eta}({}^1S_0) | 0 \big\rangle
+ \cdots \Bigg]
\nonumber\\
{\mbox{Method II:}} \;\;\;\; & \sigma(\eta) & =
\frac{\pi g^2 }{ 2 M_\eta^3}
\; \delta(E_f - E_i) \;
\bclbig 0 | {\cal O}_1^\eta ({}^1S_0) | 0 \bcrbig
\end{eqnarray}
The reader can verify that for
charmonium production from the decay of the $b$-quark,
$b \rightarrow s + J/\psi$, one obtains (for the color-singlet
contriubtions only )

\begin{eqnarray}
\label{bdecay}
{{\mbox{Method I:}}} \;\;\;\;
\sigma (J/\psi) & = &
\frac{ K }{( 24 \pi) } \; \frac{m_b^3 }{m_c} \;
\Bigg[
\big\langle {\cal O}^{J/\psi}({}^3S_1) \big\rangle
 - \frac{5}{6} \;
\frac{1}{m_c^2}
\big\langle {\cal P}^{J/\psi} ({}^3S_1) \big\rangle + \cdots \Bigg]
\\
{\mbox{Method II:}} \;\;\;\;
\sigma (J/\psi) & = &
\frac{K }{(96 \pi)  }\; \frac{m_b^3}{ m_c^2}  \;
\Bigg[
 \bclbig {\cal O}^{J/\psi}({}^3S_1) \bcrbig
-
 \frac{4}{3 m_c^2 } \bclbig {\cal P}^{J/\psi} ({}^3S_1) \bcrbig
+ \cdots \Bigg]
\end{eqnarray}
where $K$ is some constant of mass dimension $-4$ involving $G_F^2$
and CKM matrix elements.  (We have simplified things by assuming,
wrongly, that $m_c \ll m_b$.)

%

We have checked by explicit calculation that
if relativistically corrected factorization formulas for color-singlet
S-wave production are calculated for any
quarkonium production processes, in both methods, one finds that the results
can always be reconciled,
up to relative order $v^4$, using the
formulas
\begin{eqnarray}
\label{massconnectiono}
2 M_H \big\langle {\cal O}^H\big\rangle
& = &
\bclbig {\cal O}^H\bcrbig\\
\label{massconnectionp}
2 M_H \big\langle {\cal P}^H\big\rangle
& = &
\bclbig {\cal P}^H\bcrbig
\end{eqnarray}
and
\begin{equation}
\label{rcmr}
M_H = 2m_c \Bigg( 1 +
\frac{\big\langle {\cal P}^H\big\rangle  }
{2 m_c^2 \big\langle {\cal O}^H \big\rangle   } \Bigg) \big( 1 + O(v^4) \big)
=  2m_c \Bigg( 1 +
\frac{\bclbig {\cal P}^H\bcrbig  }
{2 m_c^2 \bclbig {\cal O}^H \bcrbig   } \Bigg) \big( 1 + O(v^4) \big)
\end{equation}
where $v$ is the characteristic relative velocity of the
heavy quarks in the boundstate.

The first two equations seems reasonable, since
one might think that they simply expresses
the difference between
the nonrelativistic state normalizations of Appenidix I and
the relativistic state normalizations of Appendix II.
As to the second equation, it also seems reasonable,
since it expresses the idea that relativistic corrections
to the mass of the boundstate are of relative order
$$
\frac{\big\langle {\cal P}^H \big\rangle  }
{ m_c^2 \big\langle {\cal O}^H \big\rangle   }
\sim v^2
$$

The relativistically corrected mass relation, Eq.~\ref{rcmr},
trades relativistic corrections to the mass of the quarkonium boundstate
for relativistic corrections involving the
four-fermion production matrix element
$\langle {\cal P}^H \rangle$.  Such an idea
has recently been presented in Ref.~\cite{ctone}.
These authors use the equations of motion of the heavy
quark field operators
\begin{equation}
\Big( iD_t + \frac{{\bf D}^2}{2 m_c} \Big) \psi = 0 \;\;\;\;
\Big( iD_t - \frac{{\bf D}^2}{2 m_c} \Big) \chi = 0
\end{equation}
to re-express the matrix element
$\langle {\cal P}^\eta \rangle$ so as to obtain
\begin{equation}
\label{kapustin}
\big\langle 0 | {\cal P}^\eta({}^1S_0) | 0  \big\rangle =
m_c \big( M_\eta - 2m_c \big) \;
\big\langle 0 |{\cal O}^\eta({}^1S_0)| 0 \big\rangle
\end{equation}
Solving for $M_\eta$, we find
$$
M_\eta = 2m_c \Bigg( 1 +
\frac{\big\langle {\cal P}^\eta\big\rangle  }
{2 m_c^2 \big\langle {\cal O}^\eta \big\rangle   }  \Bigg)
$$
which is exactly our relativistically correct mass relation!

\section{Differential Cross-Sections}

So far, we have considered only total production cross-sections.
For calculating differential (instead of total) cross-sections
in Method I,
one must calculate
$$\frac{d \sigma \big (c\overline{c}({\bf q}) \big)}
{d X}$$
where $X$ is a kinematic parameter such as the rapidity $y$,
the Mandelstam variable $t$, or the transverse momentum squared
$p_T^2$.   The matching condition is
\begin{equation}
\label{diffmatch}
\frac{d \sigma \big (c\overline{c}({\bf q}) \big)}{d X}
= \frac{G_n(m_c, {\bf P}, X)}{m_c^{d_n - 4}} \;
\big\langle 0 | {\cal O}_n^{c\overline{c}(q)}  | 0 \big\rangle
\end{equation}
and the factorization formula is
\begin{equation}
\label{difffact}
\frac{d \sigma \big ( H  \big)}{d X}
= \frac{G_n(m_c, {\bf P}, X)}{m_c^{d_n - 4}} \;
\big\langle 0 | {\cal O}_n^H  | 0 \big\rangle
\end{equation}
Let us contrast this situation to that in Method II.
There, the matching condition
allows one to calculate the ``effective Feynman amplitude squared''
for charmonium production, and differential cross-sections
are obtained in a straightforward manner using that object.
There is no need, in Method II, to
decide at the level of the matching which sort
of (differential) cross-section is
ultimately desired.

\section{Conclusion}

We have pointed out that there exist two different methods for calculating
factorization formulas for charmonium production, one suggested by
Ref.~\cite{bbl}, the other by Ref.~\cite{bc}.
The results of the two methods can be reconciled using
the relations
%
\begin{eqnarray}
2 M_H \big\langle {\cal O}^H\big\rangle
& = &
\bclbig {\cal O}^H\bcrbig\nonumber\\
2 M_H \big\langle {\cal P}^H\big\rangle
& = &
\bclbig {\cal P}^H\bcrbig\nonumber
\end{eqnarray}
and
$$
M_H = 2m_c \Bigg( 1 +
\frac{\big\langle {\cal P}^H\big\rangle  }
{2 m_c^2 \big\langle {\cal O}^H \big\rangle   } \Bigg) \big( 1 + O(v^4) \big)
=  2m_c \Bigg( 1 +
\frac{\bclbig {\cal P}^H\bcrbig  }
{2 m_c^2 \bclbig {\cal O}^H \bcrbig   } \Bigg) \big( 1 + O(v^4) \big)
$$

It is interesting to ask:
``which method is preferable?''  and ``Have we truly
reconciled the two methods?''  Indeed,
one might object to the step in which the equations of motion
are applied to obtain Eq.~\ref{kapustin}, since
the matrix element of the operator
$\psi^\dagger ({\bf D}^2 \chi)$ is ultraviolet divergent and requires
a subtraction that is proportional to $\psi^\dagger \chi$
\cite{thankstoeric}:
$$
\big\langle {\cal P}^\eta \big\rangle =
 m_c \big( M_\eta - 2m_c  + C \big)\;
\big\langle {\cal O}^\eta \big\rangle
$$
where $C$ is a subtraction constant that depends on the
renormalization scheme.

\acknowledgements

This work was
supported by the Robert A. Welch Foundation, by NSF Grant PHY 9511632,
and by NSERC of Canada.  The author would like to thank
Eric Braaten and Sean Fleming for helpful discussions.

\vfill\eject


\begin{appendix}

\section{NRQCD Conventions for Method I}

In performing
the matching procedure, we must evaluate the right-hand side
of the matching condition, which is written in terms of objects of
the NRQCD effective theory.  Therefore we require a set of
definitions and conventions for calculating in this theory.

Below is presented a practical set of conventions for performing
calculations in Method I. These definitions are compatible
with Ref.~\cite{bbl}, which employs the standard
nonrelativistic normalization of the quark states.
In the heavy quark state $|c({\bf p},\sigma , i )\big\rangle$,
$\sigma$ represents spin and $i$ represents color.  The states
and annihilation-creation operators are defined via
\begin{eqnarray}
\label{states}
\big\langle c({\bf p},\sigma , i) | c({\bf q},\tau , j) \big\rangle & =
& (2\pi)^3 \delta({\bf p} - {\bf q}) \delta_{\sigma\tau}\delta_{ij} \\
\big\{ a({\bf p},\sigma, i),a^\dagger({\bf q},\tau , j) \big\}  & =
& (2 \pi)^3
\delta({\bf p} - {\bf q}) \delta_{\sigma\tau}\delta_{ij} \\
a^\dagger({\bf p}, \sigma, i) | 0 \big\rangle & = &
| c({\bf p}, \sigma , i) \big\rangle \\
a({\bf p}, \sigma , i) | c({\bf q}, \tau , j) \big\rangle & = &
(2\pi)^3 \delta({\bf p} - {\bf q}) \delta_{\sigma\tau} \delta_{ij}
| 0 \big\rangle
\end{eqnarray}
The nonrelativistic heavy quark field operators are
\begin{eqnarray}
\label{fieldone}
\psi^i(x) & = & \sum_\sigma\int\frac{d^3 {\bf k}}{(2\pi)^3}
a({\bf k},\sigma , i) \xi_\sigma e^{-i k\cdot x}\\
\psi^{\dagger}_i(x) & = & \sum_\sigma\int\frac{d^3 {\bf k}}{(2\pi)^3}
a^\dagger({\bf k},\sigma , i) \xi^\dagger_\sigma e^{i k\cdot x}\\
\chi^i(x) & = & \sum_\sigma\int\frac{d^3 {\bf k}}{(2\pi)^3}
b^\dagger({\bf k},\sigma , i) \zeta_\sigma e^{i k\cdot x}\\
\label{fieldfour}
\chi^{\dagger}_i(x) & = & \sum_\sigma\int\frac{d^3 {\bf k}}{(2\pi)^3}
b({\bf k},\sigma , i) \zeta^\dagger_\sigma e^{-i k\cdot x}
\end{eqnarray}

The free-quark NRQCD operators are
\begin{eqnarray}
O_1^{c\overline{c}(q)}({}^1S_0) & = &
\chi^\dagger  \psi
       | c\overline{c}\big\rangle
 \big\langle c\overline{c}|
                  \psi^\dagger   \chi \\
P_1^{c\overline{c}(q)}({}^1S_0) & = &
\frac{1}{2} \left[
 \chi^\dagger   \psi
       | c\overline{c} \big\rangle
 \big\langle c\overline{c}|
                  \psi^\dagger
\bigg( \! \frac{i\! \stackrel{\leftrightarrow}{{\bf D}}}{2} \!\bigg)^2
 \chi \; + \; {\hbox{h.c.}} \right] \\
\end{eqnarray}
where
\begin{equation}
\label{projectionop}
|c\bar{c}\big\rangle \big\langle c\bar{c}| \;\;  \equiv \;\;
\sum_{\sigma,\tau,i,j}
|c({\bf q},\sigma ,i)\bar{c}(-{\bf q},\tau , j)\big\rangle
\big\langle c({\bf q},\sigma ,i)\bar{c}(-{\bf q},\tau , j)|
\end{equation}
It must be stressed that these operators are
intended for use in the calculation of unpolarized
production rates, {\it i.e.} production rates in which the angular momentum
of the final quarkonium state is not specified.

Using Eqs.~\ref{states} through \ref{projectionop},
one can obtain expressions for
the free-quark NRQCD matrix elements, as required by the right-hands side
of the matching condition. One obtains
\begin{eqnarray}
\label{freeone}
\big\langle 0 |
{\cal O}^{c\overline{c}(q)}({}^{1}S_0) |0 \big\rangle
& \equiv &
\big\langle 0 | \chi^\dagger \psi | c\overline{c} \big\rangle
\big\langle c\overline{c} | \psi^\dagger \chi | 0 \big\rangle
 =
N_c \sum_{\sigma \tau}
\xi^\dagger_\sigma  \zeta_\tau
\zeta^\dagger_\tau \xi_\sigma\\
\label{freetwo}
\big\langle 0 |
{\cal P}^{c\overline{c}(q)}({}^{1}S_0) |0 \big\rangle
& \equiv &
\big\langle 0 |
\frac{1}{2} \left[
 \chi^\dagger   \psi
       | c\overline{c} \big\rangle
 \big\langle c\overline{c}|
                  \psi^\dagger
\bigg( \! \frac{i\! \stackrel{\leftrightarrow}{{\bf D}}}{2} \!\bigg)^2
 \chi \; + \; {\hbox{h.c.}} \right]
| 0 \big\rangle
 =
N_c \;\; \sum_{\sigma \tau}
\xi^\dagger_\sigma  \zeta_\tau \;
\zeta^\dagger_\tau   \xi_\sigma \;\; {\bf q}^2
\end{eqnarray}
These free-quark production matrix elements appear in the
right-hand side of the matching condition given
in Eq.~\ref{bblmatchingcondo} and Eq.~\ref{bblfinal}.
Eqs.~\ref{freeone} and \ref{freetwo}
serve in passing from Eq.~\ref{cleverone}
to Eq.~\ref{happyone}.

The non-perturbative (analytically incalculable)
NRQCD matrix elements appearing
in the factorization formulas are
\begin{equation}
\big\langle 0 | {\cal O}^{\eta}({}^1S_0) | 0 \big\rangle \equiv
\big\langle 0 |\chi^\dagger  \psi
    \sum_S   | \eta + S  \big\rangle
 \big\langle \eta + S |  \psi^\dagger   \chi  | 0 \big\rangle
\end{equation}
and similarly for
$\langle 0 | {\cal P}^{\eta}({}^1S_0) | 0 \rangle $,
where $S$ indexes all soft additions to an $\eta$ particle,
such as one pion, two pions, seven pions and a photon, etc.,
with total energy less than the NRQCD cut-off.

\section{NRQCD Conventions for Method II}

Below is presented a practical set of conventions for performing
calculations in Method II.  These definitions are compatible
with the formulations in Ref.~\cite{bc}, which employs
relativistic normalization of the quark states.
\begin{eqnarray}
\label{statestwo}
\bclbig ({\bf p},\sigma , i) | c({\bf q},\tau , j) \bcrbig & =
& 2 E_q (2\pi)^3 \delta({\bf p} - {\bf q}) \\
\big\{ a({\bf p},\sigma, i),a^\dagger({bf q},\tau , j)\big\} & =
& 2 E_q (2 \pi)^3
\delta({\bf p} - {\bf q}) \\
a^\dagger({\bf p}, \sigma, i) | 0 \bcrbig & = &
| c({\bf p}, \sigma , i) \bcrbig \\
a({\bf p}, \sigma , i) | c({\bf q}, \tau , j) \bcrbig & = &
(2\pi)^3 2 E_q \delta({\bf p} - {\bf q}) | 0 \bcrbig
\end{eqnarray}
The nonrelativistic heavy quark field operators are
\begin{eqnarray}
\label{fieldfive}
\psi^i(x)&  = & \sum_\sigma\int\frac{d^3 {\bf k}}{(2\pi)^3 \sqrt{2 E_k}}
a({\bf k},\sigma , i) \xi_\sigma e^{-i k\cdot x}\\
\psi^{\dagger}_i(x) & = & \sum_\sigma\int\frac{d^3 {\bf k}}{(2\pi)^3
\sqrt{ 2 E_k}}
a^\dagger({\bf k},\sigma , i) \xi^\dagger_\sigma e^{i k\cdot x}\\
\chi^i(x) & = & \sum_\sigma\int\frac{d^3 {\bf k}}{(2\pi)^3 \sqrt{2 E_k}}
b^\dagger({\bf k},\sigma , i) \zeta_\sigma e^{i k\cdot x}\\
\label{fieldeight}
\chi^{\dagger}_i(x) & = & \sum_\sigma\int\frac{d^3 {\bf k}}{(2\pi)^3
\sqrt{2 E_k}}
b({\bf k},\sigma , i) \zeta^\dagger_\sigma e^{-i k\cdot x}
\end{eqnarray}

The free-quark NRQCD operators are
\begin{eqnarray}
O_1^{c\overline{c}(q)}({}^1S_0) & \equiv &
\chi^\dagger  \psi
       | c\overline{c}\bcrbig
 \bclbig  c\overline{c}|
                  \psi^\dagger  \chi \\
P_1^{c\overline{c}(q)}({}^1S_0) & \equiv &
\frac{1}{2} \left[
 \chi^\dagger     \psi
       | c\overline{c} \bcrbig
 \bclbig  c\overline{c}|
                  \psi^\dagger
\bigg( \! \frac{i\! \stackrel{\leftrightarrow}{{\bf D}}}{2} \!\bigg)^2
 \chi \; + \; {\hbox{h.c.}} \right] \\
\end{eqnarray}
where
\begin{equation}
\label{projectionoptwo}
|c\bar{c}\bcrbig \bclbig c\bar{c}| \;\;  \equiv \;\;
\sum_{\sigma,\tau,i,j}
|c({\bf q},\sigma ,i)\bar{c}(-{\bf q},\tau , j)\bcrbig
\bclbig c({\bf q},\sigma ,i)\bar{c}(-{\bf q},\tau , j)|
\end{equation}

Using Eqs.~\ref{statestwo} through \ref{projectionoptwo},
one can obtain expressions for
the free-quark NRQCD matrix elements, as required by the right-hands side
of the matching condition. One obtains
\begin{eqnarray}
\label{freethree}
\bclbig  0 |
{\cal O}^{c\overline{c}(q)}({}^{1}S_0) |0 \bcrbig
& \equiv &
\bclbig 0 |
 \chi^\dagger \psi | c\overline{c} \bcrbig
\bclbig  c\overline{c} | \psi^\dagger \chi | 0 \bcrbig
= 4 E_q^2 N_c \sum_{\sigma \tau}
\xi^\dagger_\sigma  \zeta_\tau
\zeta^\dagger_\tau  \xi_\sigma\\
\label{freefour}
\bclbig  0 |
{\cal P}^{c\overline{c}(q)}({}^{1}S_0) |0 \bcrbig
& \equiv &
\bclbig 0 |
\frac{1}{2} \left[
 \chi^\dagger   \psi
       | c\overline{c} \bcrbig
 \bclbig  c\overline{c}|
                  \psi^\dagger
\bigg( \! \frac{i\! \stackrel{\leftrightarrow}{{\bf D}}}{2} \!\bigg)^2
 \chi \; + \; {\hbox{h.c.}} \right]
| 0 \bcrbig
= 4 E_q^2 N_c \;\; \sum_{\sigma \tau}
\xi^\dagger_\sigma  \zeta_\tau
\zeta^\dagger_\tau  \xi_\sigma \;\; {\bf q}^2\\
\end{eqnarray}
These free-quark NRQCD matrix elements appear in the right-hand side
of the matching condition given in Eq.~\ref{bcmatchingcondo}.
Eqs.~\ref{freethree} and \ref{freefour} serve in passing from
Eq.~\ref{clevertwo} to Eq.~\ref{happytwo}.

The non-perturbative (analytically incalculable)
NRQCD matrix elements appearing
in the factorization formulas are
\begin{equation}
\bclbig  0 | {\cal O}^{\eta}({}^1S_0) | 0 \bcrbig \equiv
\bclbig 0 |\chi^\dagger  \psi
       \sum_S  | \eta + S \bcrbig
 \bclbig \eta + S |  \psi^\dagger   \chi  | 0 \bcrbig
\end{equation}
and similarly for
$\bclbig 0 | {\cal P}^{\eta}({}^1S_0) | 0 \bcrbig $.

\end{appendix}


\end{document}